\documentclass[%
 reprint,
 amsmath,amssymb,
 aps,
]{revtex4-2}

\newcommand{\tautau}{\tau^+\tau^-}

\usepackage{graphicx}
\usepackage{dcolumn}
\usepackage{bm}
\usepackage{microtype} 
\usepackage{float}
\usepackage{hyperref}

\begin{document}

\preprint{APS/123-QED}

\title{Motivation and design of a yotta-eV $\tau^+\tau^-$ collider}
\thanks{This paper is part of a semester-long research project in Siena University's {\it PHYS 400: Nuclear and Particle Physics} course.}%

\author{Matt Bellis}
 \email{mbellis@siena.edu}

\author{Matthew Carberg}
 \email{mw26carb@siena.edu}

\author{Chester Gould}
 \email{c08goul@siena.edu}

\author{Jackson Ingenito}
 \email{jc20inge@siena.edu}

\author{Fasiha Khaliq}
 \email{f05khal@siena.edu}

\author{Emely Kintzel}
 \email{ec20kint@siena.edu}

\author{Shane Kirschmann}
 \email{s23kirs@siena.edu} 

\author{Neha Matta}
 \email{nd24matt@siena.edu}
 
\author{Sophia Pavia}
 \email{sr25pavi@siena.edu}

\author{Emmett Pearl}
 \email{ea20pear@siena.edu}
 
\author{Payton Ramsdill}
 \email{pg11rams@siena.edu}

\author{Grace Scherer}
 \email{g23sche@siena.edu}

\author{Cullen Wright}
 \email{cj27wrig@siena.edu}

\affiliation{%
 Siena University
}%

\collaboration{SAINTS (Siena Accelerator Institute for Novel Tau Science) Collaboration}

\date{\today}

\begin{abstract}
Two significant goals of the particle physics community is the precision
study of the Higgs boson and the search for new particles. 
The Large Hadron Collider (LHC) is the current high-energy collider, 
soon to be superseded by the High-Luminosity LHC (HL-LHC). Much of the
community has rallied around a muon-collider, though that is most likely
25 years in the future. In this paper, we argue for a bolder approach:
{\it a tau-collider}, in which oppositely-charged $\tau$-leptons are
collided with energies on the yotta-eV scale and a potential radius
that places it in the Oort cloud. 
Given the long time-scale and significant construction
challenges, we strongly suggest the focus of the community shift
to this discovery machine. We acknowledge that the technology necessary
may require humanity to evolve to a Kardashev Level-I or Level-II civilization, which is all the more
reason to begin R\&D now.
\end{abstract}

\maketitle

\section{\label{sec:intro}Introduction}

Particle colliders have been engines of discovery for well over 50 
years. In the next few years, HL-LHC will be the primary collider
at the energy frontier over the next decade, but after that, the 
picture is uncertain. There is much excitement about a muon collider
and rightfully so. A lepton-collider provides a cleaner collision
environment than a hadron collider, and after past the past successes of
$e^+e^-$ colliders like the Stanford Linear Collider (SLC) and 
LEP (Large Electron-Positron collider), a muon collider might appear
to be the next step. In this paper, we argue that we should not 
take half-measures but rather boldly skip ahead to the final boss
of lepton colliders:
a $\tautau$ collider. The higher-mass of the $\tau$ will provide
a higher cross section for Higgs production and as a third-generation
lepton, may couple more strongly to higher-energy physics than
first- or second-generation fermions.

In the event that a 4$^{th}$ generation of fermions is discovered, 
we encourage the community to abandon this proposed effort in favour
of a collider that uses these new, higher-mass leptons. 

\begin{figure}
    \centering
    \includegraphics[width=0.95\linewidth]{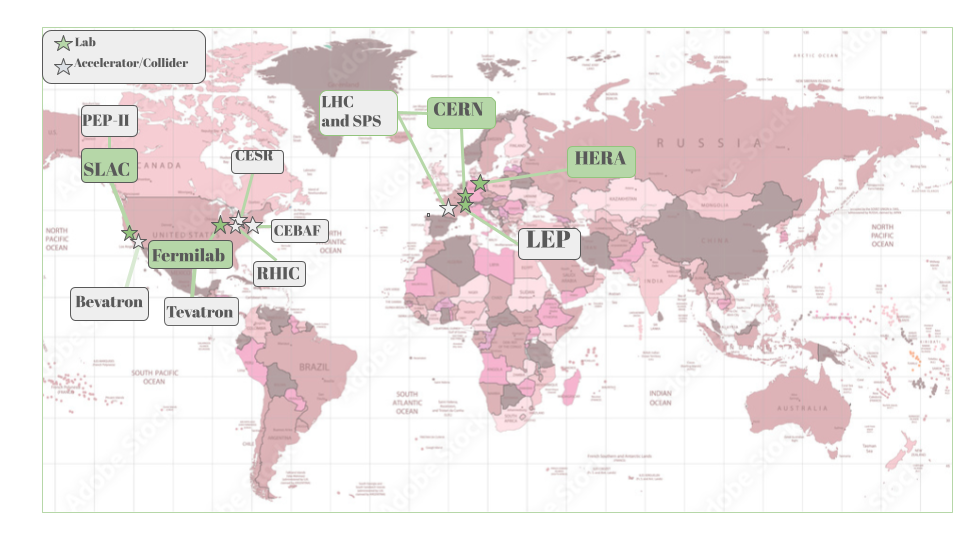}
    \caption{Map of locations of previous colliders.}
    \label{fig:colmap}
\end{figure}

\begin{figure}
    \centering
    \includegraphics[width=0.95\linewidth]{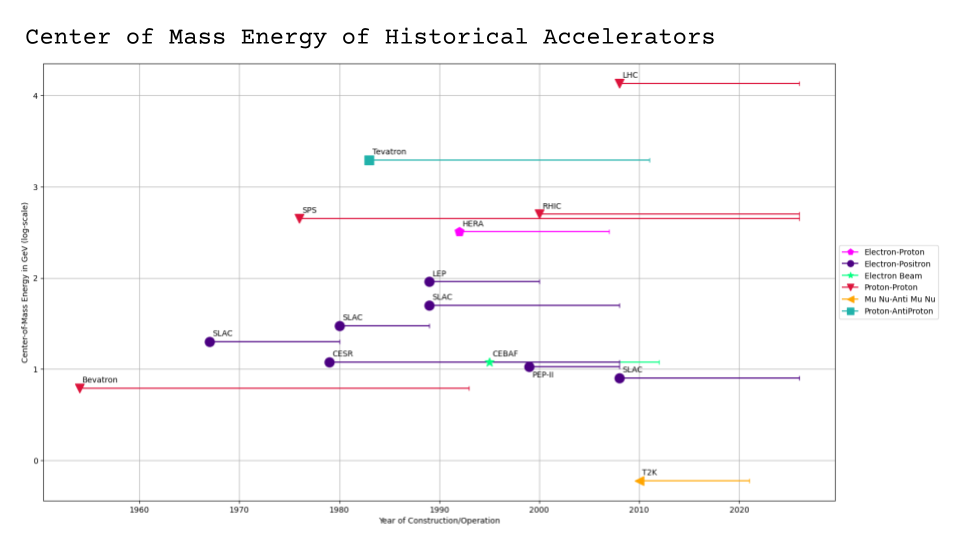}
    \caption{History of previous colliders showing years of operation and energies achieved.}
    \label{fig:colchart}
\end{figure}

Figs.~\ref{fig:colmap}-~\ref{fig:colchart} provide a graphical
representation of past and present accelerators and colliders.

\section{\label{sec:motivation}Scientific motivation}

The Standard Model consists of 17 fundamental particles which split into two groups fermions and bosons. fermions are the building blocks of the matter and these 12 fermions further split into quarks and leptons. Whereas, bosons are the force carriers and are responsible for the interactions between the matter. There are four main forces weak force carried by W and Z bosons, electromagnetic force carried by photons, strong force carried by gluons.
Gravity is not part of the Standard Model. The Standard Model cannot fully explain 
all observations, notably dark matter, dark energy, and the matter-antimatter asymmetry.
These mysteries are the reason why particle physicists are keen to discover
new physics beyond the Higgs boson. 
The Higgs boson is the last boson discovered and was predicted by Peter Higgs in 1964. 
The discovery of Higgs was important due to a burning question why do some fundamental particles have mass but others do not have?
Peter Higgs came up with the answer that an interaction with a new particle that only some particles feel must give them mass. For Higgs boson to give mass to other particles it must spread throughout the universe. All space will be filled with the invisible Higgs field made up of tiny Higgs bosons that constantly appears and disappears. The theoretical explanation of the massless photons and gluons is that they ignore Higgs field whereas, quarks, electrons and other particles interact with it so they have masses~\cite{IOP}.

\begin{figure}[H]
    \centering
    \includegraphics[width=0.5\textwidth]{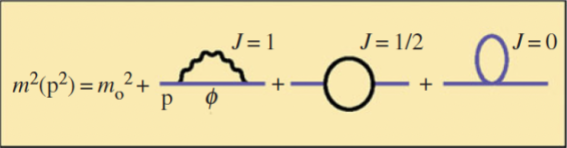}
    \caption{Radiative corrections to the mass of the Higgs boson from quantum loops containing fermions or bosons.\cite{Beyond}}
    \label{fig:radcorr}
\end{figure}

\begin{figure}[H]
    \centering
    \includegraphics[width=0.5\textwidth]{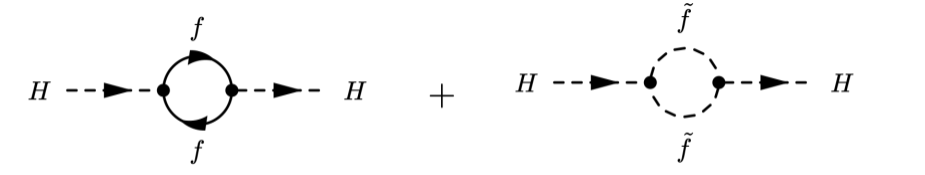}
    \caption{Supersymmetry effect on the mass of the Higgs boson.\cite{BSM}}
    \label{fig:SM}
\end{figure}

\begin{figure}[H]
    \centering
    \includegraphics[width=0.5\textwidth]{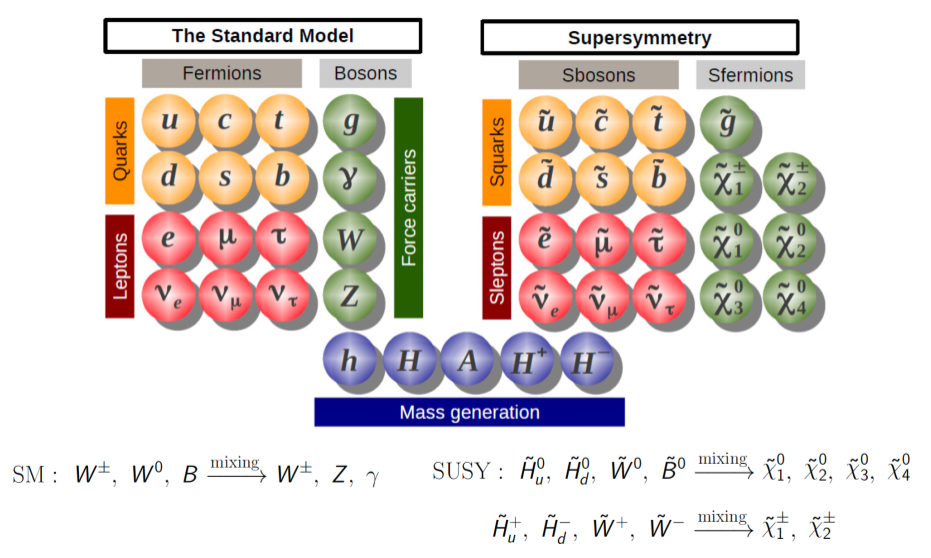}
    \caption{The Standard Model with Supersymmetric extension.\cite{BSM}}
    \label{fig:SS standard model}
\end{figure}

\textbf{Beyond Standard Model - Supersymmetry)}
There is an ample evidence of physics beyond Standard Model some of the examples are, the lightness of Higgs boson, existence of dark matter and matter-antimatter asymmetry. The mass of a fundamental scalar, such as the Higgs boson, is altered by radiative corrections.
\[
m_H^2 \to m_{\text{bare}}^2 + \Lambda^2
\]
Arising from quantum loops containing fermions and bosons.(Fig.~\ref{fig:radcorr}) where lamda is the cutt off momentum. The cutt off can be as large as $10^{15}$ GeV which means a 100 GeV Higgs will receive a huge correction. Radiative corrections have accurately predicted particle masses like the top quark and Higgs, so they cannot be ignored. This raises the key question: what protects the Higgs boson’s mass? One conjecture is that the Higgs boson is a composite object, but the current data does not support this. Another is the idea of new symmetry SUSY which says for every fermion their is a partner boson. In this way every fermion loop has a corresponding boson loop with opposite amplitude(signs) which cancels each other leaving no mass.(Fig.~\ref{fig:SM}). Due to this SUSY doubles the fundamental particle system as an example electron should have a super-partner with same mass. No such particle has ever been found so SUSY is presumed to be broken symmetry\cite{Beyond}.(see Fig.~\ref{fig:SS standard model})

\begin{figure}[H]
    \centering
    \includegraphics[width=0.5\textwidth]{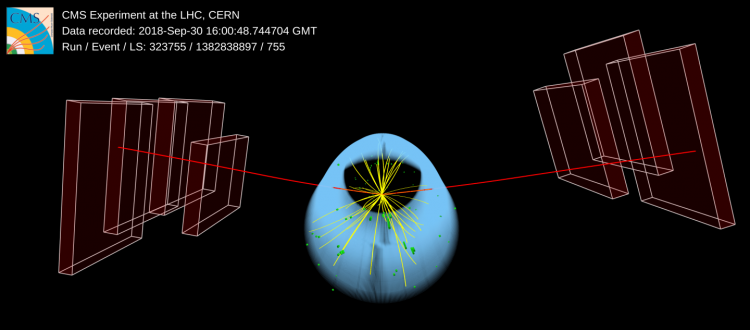}
    \caption{Higgs boson decaying to 2 muons (CMS event display).\cite{CMS}}
    \label{fig:Higgs interaction}
\end{figure}

Current colliders can only produce so many Higgs bosons and only with so much energy when we do, so having an accelerator that can collide particles that are more likely to produce Higgs bosons would be ideal. Even when we do produce a Higgs boson, they don't last very long and we are still learning about the details of their interactions. One example of such an interaction is between the Higgs and a pair of electrons.
And Higgs bosons aren't the only ones, we also don't know all we want to about top and bottom quarks. 
Another thing we don't know in sufficient detail is if the Higgs decays into particles that the Standard Model doesn't yet know about.
Some of which could be an explanation for things we theorize about but don't actually know, such as dark matter which the Standard Model cannot explain, but observation of the universe implies exists. 

\section{\label{sec:level1}Benefits of a muon collider}

There is a large fraction of the particle physics community that 
believes the next collider should be a muon collider. Here we examine
why that is. 

\textbf{Why muons?} The Standard Model is complete but universe is mainly made of dark matter(85\%) which is yet to be fully explained. To dig deeper in the dark matter particle physicists came up with the idea of a muon collider. The muon collider is more beneficial because it takes advantages of current collider technologies, though there are still engineering challenges. Muons are 200 times heavier than electrons, which means they emit 2 billion times less synchrotron radiation. Muons are fundamental particles with no internal structure unlike protons which are made up of quarks. A muon collider will use less energy than a 10 TeV muon collider and will be competitive with a 100 TeV proton collider~\cite{WinNT}.

Muons can be accelerated to the energy comparable to the protons because they are point like particles and their entire center of mass energy is available for for interactions. whereas, for protons are part of energy is available because  it is distributed statistically among partons~\cite{Muon_Collider}.

\begin{figure}[H]
    \centering
    \includegraphics[width=0.5\textwidth]{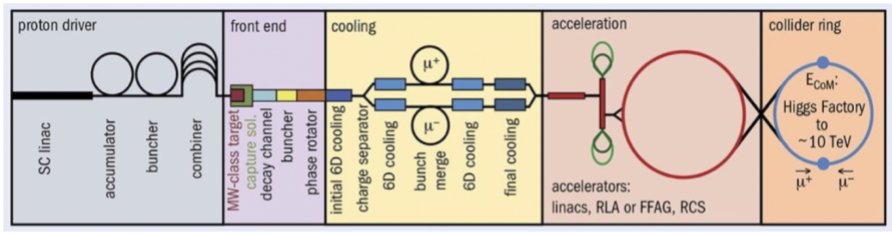}
    \caption{Muon collider conceptual design\cite{Particle_Physics}}
    \label{fig:MC Design}
\end{figure}

A muon collider will be much smaller and cheaper than a functionally equivalent proton collider. It could fit on the 2750-hectare campus of the United States dedicated particle physics lab, Fermi National Accelerator Laboratory (Fermilab)~\cite{Science} (See Fig.~\ref{fig:MC Design}.

\textbf{(Vector Boson Fusion)}. The Higgs field gives mass to fundamental particles and the observable particle in standard model is Higgs boson. By studying this physicists are searching for new phenomena's that extend beyond this and subtle deviations may appear in rare or difficult-to-measure Higgs processes.
ATLAS has improved sensitivity to Higgs decay into bottom and charm quarks. Higgs to bottom quark is 58\% of the decays and charm accounts for 3\% these processes are extremely difficult to study in hadron collider. This is due to production of collimated sprays of particles "jets" by bottom and charm quarks.  Isolating a Higgs signal amid this background is a challenging task specially during a Higgs to charm decay as charm jets are even more difficult to identify. To resolve this issue researchers focused on Higgs bosons produced via the fusion of two W or Z bosons a process called vector boson fusion(VBF). VBF  leaves a distinctive signature of two energetic forward jets\cite{ATLAS}.

Considering this a an idea of muon collider as a vector boson collider has been emerged. In this process a vector boson fusion processes will be modeled directly using VV initiated processes and the convolution of universal and process-independent parametrizations of the  emission from initial-state muons~\cite{EW_corrections}.

\textbf{(Higgs coupling to muons)}. The measurements performed so far have focused on the Higgs boson interactions with the most massive particles, such as the W and Z bosons, and only with particles from the most massive generation. The interaction of Higgs with the lighter particles in experimentally untested. measuring all the ways a Higgs boson interacts is necessary to understand whether Higgs field can explain the full range of particle masses or not. Muons are 2000 times lighter than a top quark. Higgs decaying to a muon is an extremely rare phenomena as only one is 5000 decays to a muon. The detector at CMS uses precise muon tracking using a 3.8 (Tesla) magnetic field which is about 100000 times stronger than the magnetic field of earth to measure muon trajectories with 1-2\% accuracy. This precision has allowed CMS (see Fig.~\ref{fig:Higgs interaction}) and ATLAS to report first muon and Higgs interaction ever.\cite{CMS}

\textbf{(Challenges of muons)} The first challenge to make a Muon collider is the production of the muons. The most advance area of muon collider research is proton targets. particle physicists are trying to find a robust target which can withstand the high radiation, temperatures and pressures applied by the powerful proton beam, ensuring a consistently high yield of muons. To find a solution a mercury intensive target(MERIT) experiment was performed in 2007 which demonstrated a feasible target to produce intense muon beams. Although it was a successful process to generate muon beams but mercury is hazardous to handle and susceptible to damage. Once muons are generated they can be concentrate into beams and accelerated for collision.  This requires the initially diffuse cloud of muons to be focused into a narrow, pencil-shaped beam. This process of cooling is constrained by the muon's short lifetime. When two muon beams collide at TeV energies despite the time dilation about 10\% of muons decay before collision. Their decay produces electrons, positrons and neutrinos, which can deposit energy in surrounding collider components causing radiation damage. Muon decay also produce Beam Induced Background(BIB) which can interfere in detecting decay events. particle physicists are are dedicated to addressing the many technical challenges they pose, hoping to position a muon collider as a potential successor to the High Luminosity Large Hadron Collider. IMCC and Muon Accelerator Program have collaborated to Muon detector conceptual design~\cite{Particle_Physics}.
See Fig.~\ref{fig:MC Design}.

\section{\label{sec:tau} The $\tau$ lepton}

\subsection{\label{section:tauhist} History and discovery of the $\tau$}
The search of Tau lepton was started at CERN by  Bologna–CERN–Frascati (BCF) group led by Antonino Zichichi in 1960. He suggested the idea of of a new sequential heavy lepton, now called Tau. He performed the experiment at ADONE facility in 1969 but the accelerator did not had enough energy to search Tau particle. The Tau particle was independently predicted by the Yung-su-Tsai in an article in 1971. 
\cite{WikipediA}
Given this groundwork Tau lepton was detected by Martin Lewis Perl in 1975 at SLAC using MARK1 detector at SPEAR(Stanford Positron Electron Asymmetric Rings) accelerator. It consists of 80m(diameter) ring. Tau lepton discovery was at center of mass energy of 4.8(GeV). \cite{tau}
At SLAC they were not able to discover Tau directly rather they discovered unusual events. They discovered 64 events of the form,
\[
e^+ + e^- \rightarrow e^{\pm} + \mu^{\mp}
\]
They had at least 2 undetected particles. These two undetected particles were needed because energy and momentum cannot be conserved with just one. There was no detection of muon, electrons or any other hadrons. Which led to the thought of production and decay of a new particle pair.
\[
e^+ + e^- \to \tau^+ + \tau^- \to e^\pm + \mu^\mp + 4\nu
\]
For this discovery Martin Lewis Perl was awarded with the Nobel prize in 1995~\cite{WikipediA}.

The tau particle has a mass of 1776.93 $\pm$ 0.09 MeV, which is larger than even some hadrons. This larger mass is the reason why it's llifetime is shorter than the lighter leptons. This is why the tau particle's lifetime is so problematically short(290 $\pm$ 0.5 femtoseconds)~\cite{PDG} if we want to to use the taus in a $\tau^\pm$ collider.

\section{\label{sec:motivation}A $\tautau$ collider}

There are many design choices that must be made that are driven
by the scientific goals and engineering limitations. Here we explore
the challenges and opportunities ahead. 

\subsection{\label{sec:lincirc} Linear vs.Circular Accelerator}
 One of the main design choices for a collider is whether to use a linear or circular geometry. Circular colliders use magnetic fields to bend charged particles into closed loops, allowing the same particles to collide many times thus increasing the overall collision rate. However, this design becomes inefficient at very high energies due to synchrotron radiation. When charged particles move along curved paths, they emit energy, and the power radiated scales as $P \propto \gamma^4 / R^2$, where $\gamma$ is the Lorentz factor and $R$ is the radius of the accelerator \cite{wiedemann}. Because of the strong dependence on $\gamma$, energy losses increase rapidly at high energies, making it extremely difficult to maintain the beam in a circular collider. This effect is especially significant for leptons such as the tau, which lose much more energy to radiation as they are heavier particles \cite{fcc}.

Linear colliders avoid this issue because particles travel in straight lines and do not undergo centripetal acceleration, meaning there are no synchrotron radiation losses from bending. This allows linear designs to reach much higher energies, which is essential for a $\tau^+ \tau^-$ collider. In addition, the short lifetime of the tau lepton limits the efficiency of storing particles for repeated collisions, which is the main advantage of circular colliders \cite{pdg2022}. Even with relativistic time dilation, taus decay quickly; the benefit of multiple passes is minimal and next to impossible. For these reasons, a linear collider is the most practical choice, as it avoids major energy losses and is better suited to achieve the extreme conditions required for a $\tau^+ \tau^-$ collider. 

\subsection{\label{sec:level2} $\tau$ lifetime and relativistic effects}

Our collider needs to accelerate and focus the $\tau$'s
before colliding them and so we need to accelerate them 
to energies such that relativistic time dilation will 
allow more of them to survive before decaying. 

{\bf Student analysis \#1}. The tau lepton has a very short proper lifetime of $\tau_0 = 2.9 \times 10^{-13}\text{s}$ and a rest mass energy of $m_\tau c^2 = 1.777\text{GeV}$ \cite{pdg2022}. In order to make taus usable in a collider, their lifetime in the laboratory frame must be extended through relativistic time dilation, given by $\tau_{\text{lab}} = \gamma \tau_0$, where $\gamma$ is the Lorentz factor. Solving for $\gamma$, we obtain $\gamma = \tau_{\text{lab}} / \tau_0$. For a desired lifetime of $1,\text{ms}$, this gives $\gamma \approx 3.45 \times 10^{9}$. The corresponding energy can be found using $E = \gamma m_\tau c^2$, which yields $E\approx 6.1 \times 10^{18}\text{eV}$. 

For a much longer lifetime of $1,\text{s}$, the required Lorentz factor increases to $\gamma \approx 3.45 \times 10^{12}$, leading to an energy of $E\approx 6.1 \times 10^{21}\text{eV}$. These energies are many orders of magnitude beyond what current accelerators can achieve, as modern facilities operate near $10^{4},\text{GeV}$. For example, the LHC operates at 13,600 GeV, and the Tevatron operated at 1960 GeV. This demonstrates that even modest increases in the $\tau$ lifetime require extremely large relativistic effects, making it very challenging to produce and sustain $\tau$ beams for meaningful experimental use~\cite{pdg2022}.

{\bf Student analysis \#2}. To have a tau live for 1 ms, we will use the time dilation formula $t = \gamma t'$ to solve for the relativistic factor needed to make the tau live for 1 ms. We can then use this relativistic factor to find the speed the tau needs to travel at to live for 1 ms in the lab frame. We use the value for the lifetime of the tau from the Particle Data Group~\cite{pdg2022}.

$$ \gamma = \frac{t}{t'} = \frac{1\times 10^{-3}s}{2.9\times 10^{-15}s} = 3.448\times10^{11}$$

This is an astronomically large value for gamma, and in order to figure out what speed the tau particle needs to be accelerated to, we would use 
$$\beta = \sqrt{1-\frac{1}{\gamma^2}} = \sqrt{1-\frac{1}{(3.448\times10^{11})^2}} = 0.99...$$ 

With 23 digits of 9 in total. This is 99.99999999999999999999957$\%$ the speed of light. We can already tell that the energy needed for each particle to reach these speeds is beyond anything that we can produce on earth today. To find the energies needed for each particle, we can use:
$$E = \gamma mc^2 = (3.448\times10^{11})(1.776 [GeV])$$
$$= 6.12\times10^{11} [GeV] = 344.8[EeV]$$
Each tau beam would require an energy of 344.8 Exa electron volts, an impossible jump from today's LHC energy of 6.5 TeV per beam. Even at this energy, the tau beam would only live for 1 ms! to have the tau beam survive through multiple passes of a large circular collider, it would need to live on the order of minutes or even hours. For the taus to live for 10,000 seconds, or 2.7 hours, the relativistic factor would need to be on the order of $3.448\times10^{18}$, and would require an energy per beam of $E = 6124[YeV]$, or 6124 Yotta electron volts.

{\bf Student analysis \#3}.
With such a short lifetime, most tau leptons travel around the length of a nucleus. While creating a tau collider seems impossible it can be done just at very high energies thanks to relativity. As we can use the relativistic factor of $\gamma = 3.45 \cdot10^9 $ and find the momentum needed to make a tau live around 1 ms.
Using the equation $ p = \gamma m\beta c$ we can find the momenta needed.
\begin{table}[h!]
\centering
\begin{tabular}{|c|c|}
\hline
\textbf{Quantity} & \textbf{Value (GeV/c)} \\
\hline
$P_t$ & $1.8 \times 10^{18}$ \\
\hline
$P_z$ & $9.4 \times 10^{14}$ \\
\hline
$P_e$ (each) & $5.2 \times 10^{17}$ \\
\hline
Total $P_t$ (full collision) & $2.5 \times 10^{18}$ \\
\hline
\end{tabular}
\caption{Momentum required to create a tau with a lifetime of 1 ms}
\end{table}

By also trying the relativistic equations for various energies we found that an energy with magnitude of 10 zetta eV, shown in the table below.

\begin{table}[h!]
\centering
\begin{tabular}{|c|c|c|c|}
\hline
Particle & $E$ (GeV) & $\gamma$ & Lifetime (s) \\
\hline
\multicolumn{4}{|c|}{Muon} \\
\hline
$\mu$ & $1.00 \times 10^{3}$  & $9.46 \times 10^{3}$  & $2.08 \times 10^{-2}$ \\
$\mu$ & $1.00 \times 10^{4}$  & $9.46 \times 10^{4}$  & $2.08 \times 10^{-1}$ \\
$\mu$ & $6.00 \times 10^{6}$  & $5.68 \times 10^{7}$  & $1.25 \times 10^{2}$  \\
$\mu$ & $1.00 \times 10^{7}$  & $9.46 \times 10^{7}$  & $2.08 \times 10^{2}$  \\
$\mu$ & $1.00 \times 10^{10}$ & $9.46 \times 10^{10}$ & $2.08 \times 10^{5}$  \\
$\mu$ & $1.00 \times 10^{13}$ & $9.46 \times 10^{13}$ & $2.08 \times 10^{8}$  \\
$\mu$ & $1.00 \times 10^{14}$ & $9.46 \times 10^{14}$ & $2.08 \times 10^{9}$  \\
\hline
\multicolumn{4}{|c|}{Tau} \\
\hline
$\tau$ & $1.00 \times 10^{3}$  & $5.63 \times 10^{2}$  & $1.63 \times 10^{-12}$ \\
$\tau$ & $1.00 \times 10^{4}$  & $5.63 \times 10^{3}$  & $1.63 \times 10^{-11}$ \\
$\tau$ & $6.00 \times 10^{6}$  & $3.38 \times 10^{6}$  & $9.79 \times 10^{-9}$  \\
$\tau$ & $1.00 \times 10^{7}$  & $5.63 \times 10^{6}$  & $1.63 \times 10^{-8}$  \\
$\tau$ & $1.00 \times 10^{10}$ & $5.63 \times 10^{9}$  & $1.63 \times 10^{-5}$  \\
$\tau$ & $1.00 \times 10^{13}$ & $5.63 \times 10^{12}$ & $1.63 \times 10^{-2}$  \\
$\tau$ & $1.00 \times 10^{14}$ & $5.63 \times 10^{13}$ & $1.63 \times 10^{-1}$  \\
\hline
\end{tabular}
\caption{Relativistic gamma factors and lifetimes for muons and tau leptons at various energies.}
\end{table}
\cite{PDG}

\subsection{\label{sec:eng}Engineering challenges}

We will need high-gradient electric fields in order to accelerate
the $\tau$'s. Here we explore the challenges of both accelerating
them and bending them in our accelerator.

\subsubsection{\label{sec:elec}Electric fields}

In order to accelerate the $\tau$ to speeds which can result in a high energy collision we must be able to generate a high energy Electric field. In the LHC, along with most other colliders today, field generation is accomplished with RF cavities, a large chamber held at supercooled temperatures inside of which RF electromagnetic waves resonate, forming a field which can impart up to 2 million volts of energy per meter of length~\cite{LHC_RF_Cavities}. 

It is important to note that for the purposes of making these calculations easier and more legible, all calculations will be made for energy, rather than momentum. That being said, for high energies, the mass will have a negligent effect on the energy, so E will be essentially equal to $\rho$ for all practical purposes, and given the scale we will be working with this will be true a vast majority of the time. For any particle of charge e (the standard charge of an electron) the electric field can be thought to impart $1 \frac{MeV}{C}$ of energy for every $MV$ of potential the particle passes through. This means that the RF cavity would impart 2 MeV of energy for every meter it passes through. We will be using these 2 rules as the basis for all of our calculations.

Understanding the magnitude of the task before us, it should be clear that we will need far more energy than $2 \frac{MV}{m}$, which leads us to looking at alternative energy sources. The most realistic alternate energy source is the Plasma Wakefield. Plasma Wakefield is a developing electric field production method which relies on shooting a laser through a plasma. The laser then causes the loose electrons in the plasma to begin moving, this motion leaves a "wake", called so because it resembles the wake of a boat traveling along water \cite{CERN_wakefield}. This wake is capable of reaching extremely high energies, with current predictions hoping to get up to $10 \frac{GV}{m}$ \cite{plasma_acceleration}. Even so, there are several challenges with this method, including dephasing, pump depletion, and beam diffraction, which all pose massive challenges to real world implementation.

Lets pretend for a moment though that we found an electric field generation method which was more stable, and stronger than the Plasma Wakefield technology. The largest electric field possible, thanks to the Schwinger effect, would be nearly $1.32 \times 10^{18} \frac{V}{m}$ \cite{Schwinger_limit} which is important because this means that a field capable of accelerating our $\tau$ to its ideal energy of around $1 \times 10^{22} \frac{eV}{c}$ is theoretically possible. Unfortunately, the largest man-made electric field is a small fraction of that limit, only reaching into the $TV$ range, up to $83 \frac{TV}{m}$ for a brief second at a lab in japan by firing a high powered laser at a piece of silver foil \cite{Large_E_Field}.

These energies were reached only very briefly, but for the purposes of exploration lets assume that we could create an electric field that is at that same power, and is stable enough to be used in a real-world collider. In order to reach our desired energy of $1 \times 10^{22} \frac{eV}{c}$ the particle would have to move through at least $1.2*10^9 m$ of an E field at a minimum energy of $83 \frac{TV}{m}$. Knowing how short our life-time is, lets assume we wanted to accelerate the $\tau$ to its ideal speed in only 1 km of field exposure. Under these conditions, the field would need to be $100 \frac{EV}{m}$. If we were instead using the more realistically achievable $10 \frac{GV}{m}$, we would need to travel through $1 \times 10^{13} m$.

\subsubsection{\label{sec:level1} Magnetic fields}

Magnetic fields are generated by electric currents and therefore bound to the flow of electric current\cite{B_Field_strength}. Therefore, the strength of the magnetic field is bound to how strong the current is \cite{B_Field_strength}. The highest velocity reachable is the speed of light which means that our magnetic field is limited by

$$\nabla \times B <\mu_0 e Nc$$

Therefore our magnetic field can reach a maximum strength of $6\times10^{-8}$ $N_{cc}$ $L_{km}$ \cite{B_Field_strength}. "Here $N_{cc}$is in units of electrons per $cm^{-3}$ and $L_{km}$ is the length scale across a current filament in units of km" \cite{B_Field_strength}. But this leaves a question open. How can magnetars or other compact objects, which do not have dynamo action, build and sustain their magnetic field and does it have the same upper limit? These fields are produced through differentially rotating progenitors~\cite{B_Field_strength}. This means that there are different layers which rotate at different angular velocities. Neutron stars then produce their strong magnetic fields when the heavy progenitor star collapses and then doesn't have the time to dissipate the magnetic energy~\cite{B_Field_strength}. This highers the level for possible magnetic strength to  $B \lesssim 10^{35}$ Gauss because we have compression factor of $\sim10^{12}$~\cite{B_Field_strength}. By decreasing the scale of heavier objects like neutron stars or magnetars could reach much stronger magnetic fields~\cite{B_Field_strength}. However this is limited by that such objects need to become black holes when collapsing, according to the no-hair theory~\cite{B_Field_strength}.

Designing our accelerator, we need to play close attention to how strong our magnets need to be. This will depend on the radius of our accelerator as well as on the Energy of the taus~\cite{Intro_accel}. 
Magnetic rigidity is a measure for how much magnetic field strength is required for bending a particles path~\cite{Intro_accel}. We can derive the equation by setting the Lorentz force equal to the Centripetal force\cite{Intro_accel}. 

$$F_L = F_C$$
$$evB = \frac{mv^2}{\rho}$$

Now we solve for magnetic rigidity by solving for the magnetic field strength multiplied by the local bending radius~\cite{Intro_accel}. 

$$B\rho=\frac{mv}{e} =\frac{p}{e}$$
We can then manipulate the equation using: $$p=\frac{\beta E_{tot}}{c}$$ This gives us our final equation for magnetic rigidity~\cite{Intro_accel}.

$$B\rho=\frac{\beta E_{tot}}{ce}$$

\begin{figure}[H]
    \centering
    \includegraphics[width=0.5\textwidth]{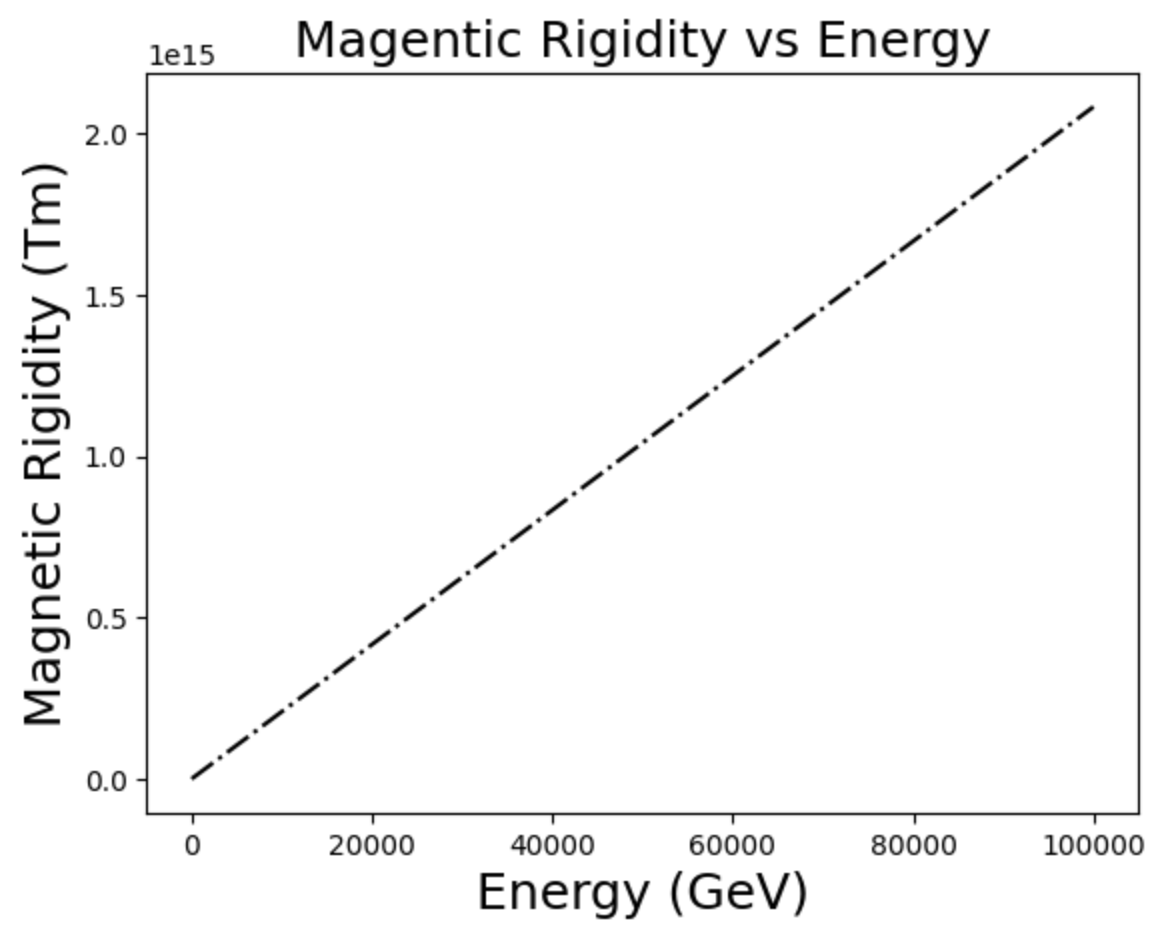}
    \caption{This graph shows a linear relationship between magnetic rigidity and the energy of tau particles.}
    \label{fig:magridg}
\end{figure}

Fig.~\ref{fig:magridg} shows that the magnetic rigidity is proportional to the energy of tau particles. Therefore if we increase the energy of the tau, then its magnetic rigidity will increase as well. This means that when we build our accelerator and use high-energy taus then we will need stronger magnetic fields because the higher the energy the harder they become to bend. From this graph we can see that we need to design our magnets according to the energy of our taus. 

\begin{figure}[H]
    \centering
    \includegraphics[width=0.5\textwidth]{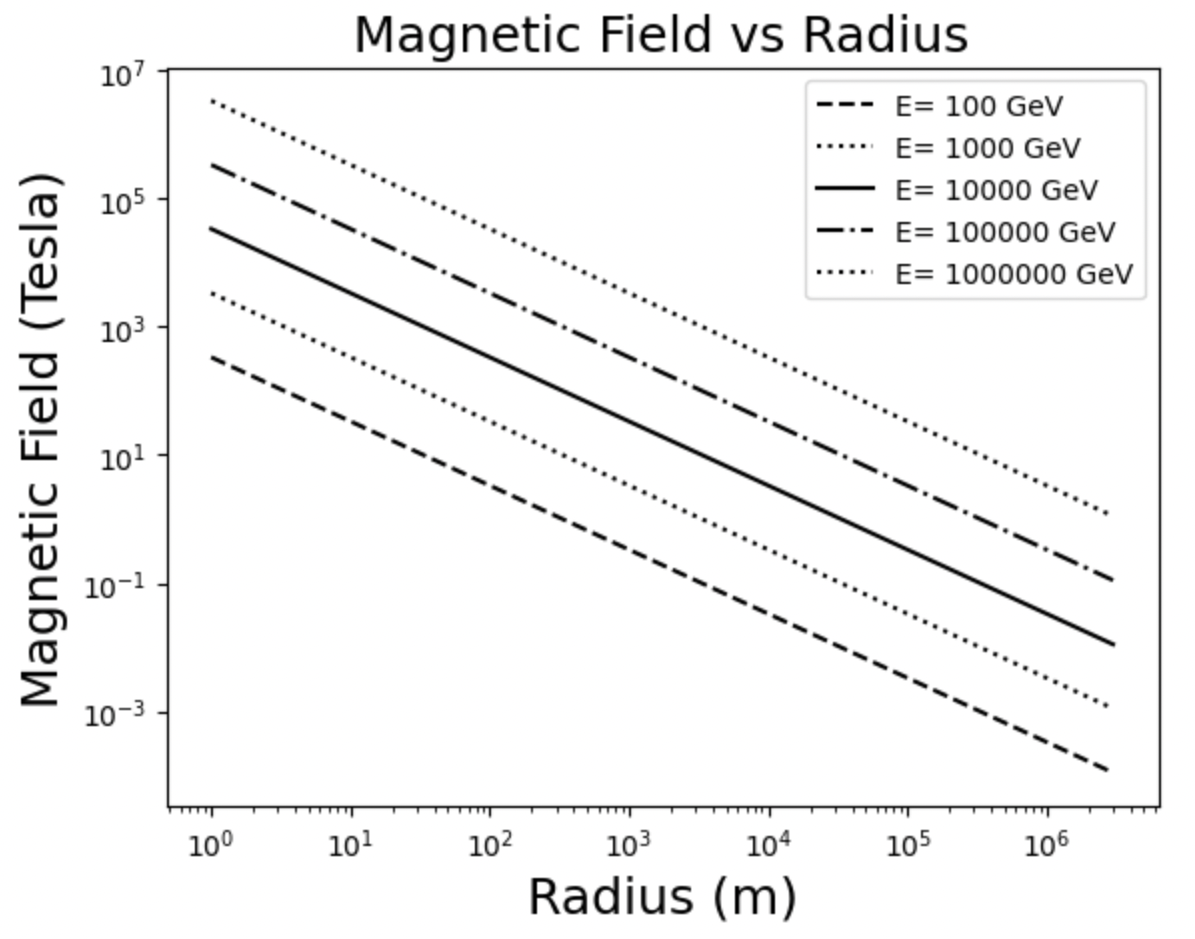}
    \caption{This graph plots the relationship of the magnetic field strength with respect to a varying radius for different energies.}
    \label{fig:magrad}
\end{figure}

\begin{figure}[H]
    \centering
    \includegraphics[width=0.5\textwidth]{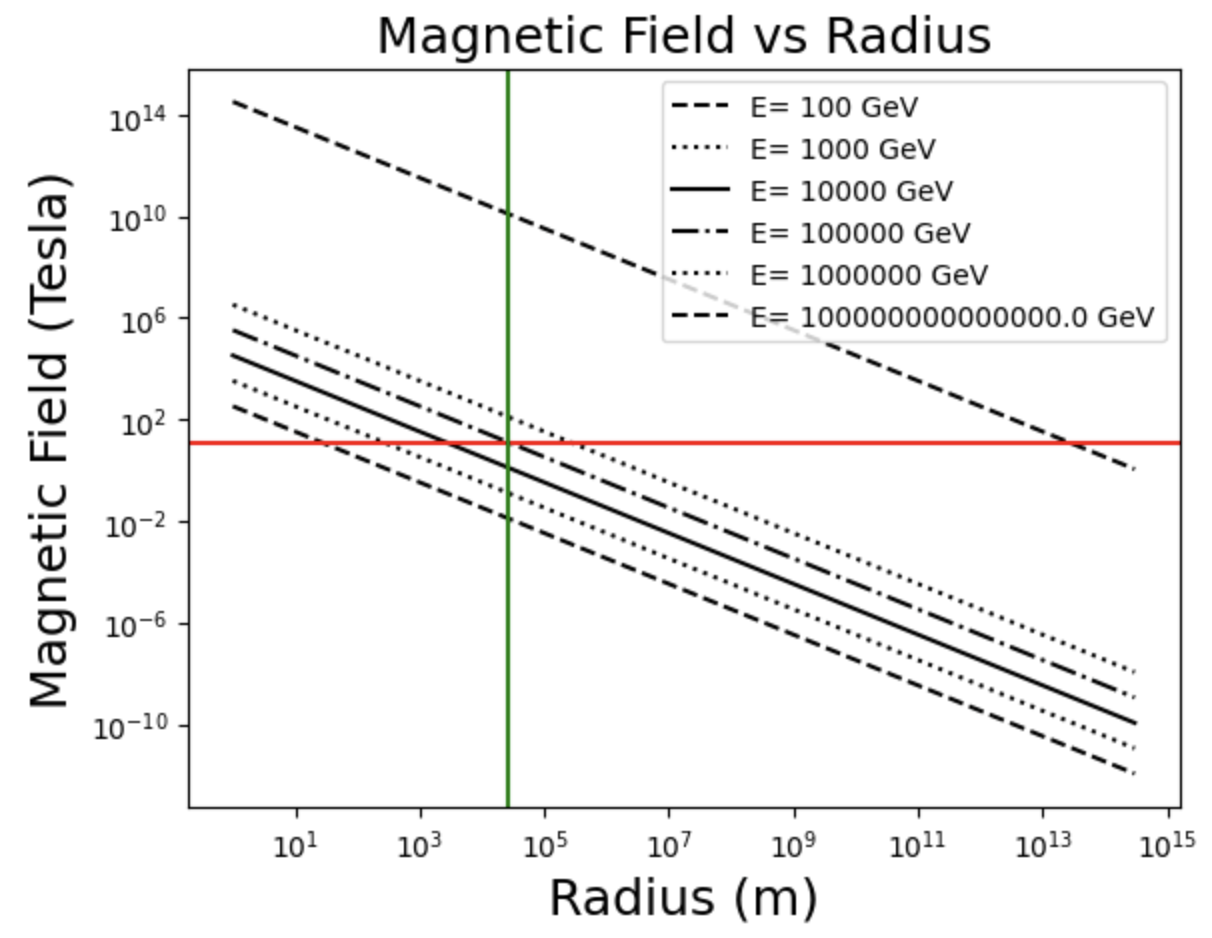}
    \caption{This graph plots the relationship of the magnetic field strength with respect to a varying radius for different energies. In addition we have a vertical line for a 27000 m radius which was our aimed radius and a horizontal line at 11 T. In addition we added our new calculated energy of Zeta electron Volts}
    \label{fig:1}
\end{figure}

In Fig.~\ref{fig:magrad} we see what magnetic field strength (Tesla) is required based on different radii (meters) on a log-log scale. The graph explored this relationship for four different energies. Starting at 100 GeV up until 100000 GeV. Given this we can see that the lower the Energy of the taus is the higher the Magnetic Field strength needs to be for the same radius. Similarly if we keep the same energy but change the radius of the accelerator we need less magnetic strength if the radius is bigger. This shows an inverse relationship between the magnetic field strength and radius. This is significant for the design of our accelerator because if we increase its radius we can use a weaker magnetic field. In figure 3 we added a horizontal line at 11 Tesla and a vertical line at 27000 meters. In our first design we wanted to use a radius of 27000 m so if we follow the vertical line and look at the intersection we can see how much magnetic field strength is required for each energy given in the plot.

Why are we using superconducting magnets? Other options would include Normal-conducting iron-dominated magnets which are limited by iron saturation which limit the Magnetic flux density down to two Tesla~\cite{superconduction_CERN}. When a ferromagnetic object is magnetized its magnetic force initially increases in proportion to the magnetizing field~\cite{Mag_Saturation}. After a certain point however saturation is reached and the magnetic force barely increases~\cite{Mag_Saturation}. Due to magnetic saturation we can not have unlimited strong magnets~\cite{Mag_Saturation}. 
Superconducting magnets, on the other hand, are limited by only the material properties~\cite{superconduction_CERN}. Superconducting materials were first discovered in the year 1911 by Kammerlingh-Onnes, who found zero resistance of a mercury wire at 4.2 K~\cite{superconduction_CERN}. The temperature at which this takes place is known as critical Temperature\cite{superconduction_CERN}. It is interesting to note that superconductivity is not found in the most conductive materials but in lots of different ones~\cite{superconduction_CERN}. Two years later he proposed a solenoid which would be able to endure 10 T, but it took fifty years to realize this~\cite{superconduction_CERN}.
Superconductivity solves problems like loosing Energy in form of heat through resistance, since every material has some sort of resistance towards the movement of electric resistance~\cite{DOE_superconductivity}. Usually this resistance stays consistent even at low temperatures, except for superconductive materials~\cite{DOE_superconductivity}. Superconductivity is the property of a material to conduct direct current without losing energy when they are cooled below their critical temperature~\cite{DOE_superconductivity}. Due to that, there is no magnetic field within the superconductor but around it, using it as almost a perfect dipole~\cite{DOE_superconductivity}.


\subsection{\label{sec:level1}Dangers? The Ring of Death}
Decaying leptons emit neutrinos that travel horizontally through the Earth. These neutrinos can interact with atomic nuclei in the ground, producing secondary particle showers that may emerge as dangerous radiation at the surface. This is one of the main challenges particle physicists encounter when designing a high energy lepton collider, often referred to as the ``ring of death''. In general, this radiation is not a concern for taus because they decay almost instantly. However, if they were to live for 1 ms they could be focused into a circulating beam, which would produce a neutrino radiation hazard.

$\tau$ decays can produce more neutrinos than $\mu$ decays due to the large number of available branching fractions. While $\mu$ decay primarily into a $\nu_{\mu}$ and an $\bar{\nu}_e$, $\tau$ have many possible decay modes, many of which produce multiple neutrinos in the final state. As a result a $\tau$ beam would generate a higher neutrino flux than a $\mu$ beam, leading to a higher radiation dose.

Using the equation shown in \cite{Bevelacqua2012}, the radiation produced by a hypothetical lepton collider can be estimated using
\begin{equation}
H_0 = K\frac{N}{g} \int_0^{E_0 }E \Omega(E) dE
\end{equation}
where $N$ is the number of lepton decays, $g$ is the accelerator gradient, and $\Omega(E)$ accounts for the neutrino cross section and tau decay effects. This equation shows that the radiation dose increases with energy, since increasing the upper limit of the integral ($E_0$) increases the value of the integral.

If $\Omega(E)$ is assumed to be stable, the integral can be evaluated as
\begin{equation}
\int_0^{E_0} E\, dE = \frac{E_0^2}{2}
\end{equation}

which implies that the radiation dose scales approximately as $H \propto E^2$. Thus, the dose at a higher energy can be estimated using

\begin{equation}
\frac{H_2}{H_1} = \left(\frac{E_2}{E_1}\right)^2
\end{equation}

According to the values shown in \cite{Bevelacqua2012}, a $\tau$ collider operating at a beam energy of 5 PeV produces a radiation dose of approximately $3.4 \times 10^4$ mSv/year, these values serve as $E_1$ and $H_1$, respectively. For taus to survive for 1 ms, a beam energy of approximately 6 PeV is required (see Part III), which serves as $E_2$. Using Eqn.~(3) and solving for $H_2$ gives

\begin{equation}
H_2 = (3.4 \times 10^4)\left(\frac{6000}{5000}\right)^2
\end{equation}

This corresponds to approximately $4.9 \times 10^4$ mSv/year, which is significantly above lethal radiation levels. Therefore, if $\tau$ living for 1 ms were condensed into a collider beam, the resulting radiation would be extremely large.

While constructing the collider tunnel deep underground or building the collider on movable mounts to adjust its orientation could reduce the radiation \cite{Muon_Shot}, these approaches do not eliminate the problem. This radiation could pose serious risks to human life and the surrounding environment. As a result, a tau collider would be extremely difficult to realize with current technology.


\subsection{\label{sec:create}Producing the $\tau$ leptons}

Before the $\tau$'s can be collided, they first must be produced
through secondary processes, after which they can be steered
out from the primary beam collision point. Here we explore methods currently
used to produce secondary particle beams and then extrapolate to new
approaches.

\subsubsection{\label{sec:neutrinobeams} Neutrino Beams}

To create taus, it is possible to collide protons to a fixed target that creates a high level of particles, and then isolate the secondary particles that are produced from the interaction. The design and understanding of this can be obtained by studying the production of neutrino beams. To produce a neutrino beam, protons collide with a heavy nuclear target that produces pions and kaons. These particles decay into neutrinos that are then focused to produce neutrino beams. The neutrino beams are used to study the weak and electromagnetic interactions. For example, at the Deep Underground Neutrino Experiment (DUNE), protons collide with a graphite target which produces neutrons, $\pi^+$s, and $\pi^-$s. Because of the energy of the collision, particles are moving in a variety of different angles and the spread of this beam is very large. A magnetic horn is used to select positive pions, and focus them into a narrow beam. This device and it's use is further explored in the section below. The positive pions spontaneously decay into positive muons and neutrinos, and concrete and steel are used to slow and absorb the muons. Neutrinos are able to pass through these materials, leaving us with a neutrino beam that we can use to study rare interactions. Rare interactions can also be studied using tau collisions. 

\begin{figure}[H]
    \centering
    \includegraphics[width=0.5\textwidth]{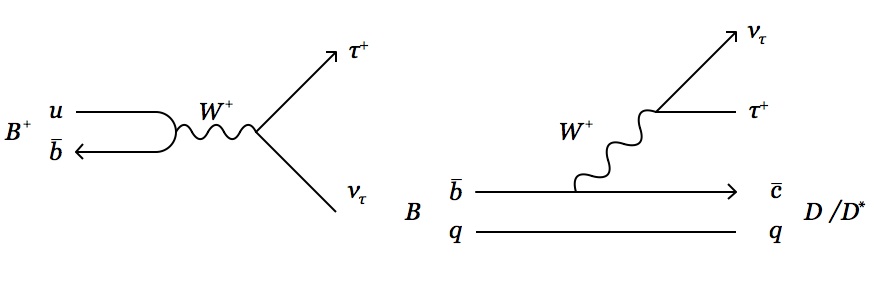}
    \caption{Feynman Diagrams of the decay modes of a B meson with taus in the final state.  Branching ratios: $1.09 \pm 0.24 \times 10^{-4}$ (left) and $7.7 \pm 2.5 \times 10^{-3}$ (right). \cite{Quantum_Diaries}}
    \label{fig:5}
\end{figure}

Studying the tau itself would not justify the immense amount of resources and energy required to build this collider. However, because of the $\tau$'s heavy mass, $\tau$ collisions have the potential to create the Higgs Boson, a particle that we have an incomplete understanding of. As mentioned above, a potential way to create taus is to facilitate a collision with a high center of mass to create particles that decay into taus. One such particle is the D meson, which can be created using a proton and heavy ion collider, and decays into the tau and tau neutrino. The branching ratio of this kind of decay is $2.1 \times 10^{-3}$, meaning it is rare. Another option is the B meson, which has multiple decay modes that include taus in the final state. The Feynman diagrams and branching ratios of these modes are shown in Fig.~\ref{fig:5}. This interaction has only been studied at an electron-positron collider. ~\cite{Accelerator_Neutrino_Beams, B_meson_decays, How_to_make_a_Neutrino_Beam, PDG}

There are several methods one might consider to focus secondary beams before they decay into neutrinos, one of which is something called "magnetic horns" that use magnetic fields to focus all particles of one sign into a narrower beam as they pass through it, and de-focus all particles of the opposite sign causing them to scatter.
A parabolic horn works for all angles of entry(within certain degrees of precision), and the quality of focusing is independent of momentum BUT the focal length is, it corrects all particles with the same momentum, charge and lifetime by the same angle, but the focal length changes depending on these variables, depending the most on momentum. Ellipsoidal ones seem to be basically parabolic horns but they work better to a certain degree. Really the perfect lens is one that is a compromise between a true parabola and a true ellipse. Another form of  horns are "magnetic fingers" are an improvement upon the improvement, named that way because the shape it takes reminded the designer of it of a finger. The cross section looks like a cone that then becomes an ellipse with the particles coming through the point of the cone and hitting the target in the middle of the end of the ellipse. Generally It's strategic to use multiple lenses, sometimes attached and sometimes separated by a great distance. The basic idea is that particles that don't get focused by the first enough pass through the second and get focused again, then those get focused again by a third lens if deemed necessary, etc.
One of the other best methods are quadrupoles, which are generally less efficient than horns but are cheaper to produce, easier to design, and provide exact focusing for particles with a specific P. Having 1 quadrupole will focus the beam in one plane but defocus it in another, having a second quadrupole with perpendicular magnetic fields will solve this issue, and having a third quadrupole will make a larger difference than you might think because even though having two does technically cover every angle, having three keeps it contained more tightly. Horns focus one sign while defocusing the other, however quadrupoles focus both signs equally, which means that if you want to filter out one sign of particles you would need a dipole upstream. The third method(albeit a less used one) are plasma lenses. The basic principle is that the beam goes through a cylinder of Plasma and the plasma produces a magnetic field similar to a horn, but this field can be adjusted more easily in the plasma than in a horn, and the beam(if I understand correctly, I'll have to fact check) would have to go through the ends of the cylinder containing the plasma, which is generally too impractical to make plasma lensing an option. And finally for the fourth method, DC current producing magnetic fields have also been considered, mainly in the form of A: fins with wires in them that produce the fields and B: having the beam pass through a solenoid to focus it and accelerate it. Narrow band beams are often desirable because they filter out less desirable particles by splitting the beam into separate bands, and having the undesirable bands be redirected into decay tunnels. These are called dichromatic beams not because of color(I think) but because of the 2 bands that are formed as opposed to having one band in a monochromatic beam. Sometimes these beams have plugs made out of heavy elements like tungsten that are placed such that the undesirable particles that decay into neutrinos later are blocked by the lead. Some have attempted to assist in the production of narrow band beams by using dipoles upstream of a horn, however in practice the dipole can interfere with the horn's efficiency a little. Off axis neutrino beams are something to be aware of because any particles that scatter may be assumed to continue traveling in a direction more or less parallel to the axis, however the further from the axis they are the harder they are to focus so there are bands of particles that decay into neutrinos traveling more or less parallel to and around the main beam that's being focused and accelerated.

\subsection{\label{sec:preferred} Preferred $\tau$-production approach: asymmetric $Z$-boson production}

Here we lay out a proposal to build a highly asymmetric $e^+e^-$ collider
that produces $Z$ bosons which decay to $\tautau$ pairs, which can
then be steered into our larger collider. There are historical
examples of asymmetric colliders being used for specific physics
goals.

\subsubsection{\label{sec:pepii} PEP-II}
To begin this section, we need to discuss relevancy. The PEP-II's was the collider used by the BaBar experiemtn at SLAC. Its main goal was to create a time-dependent measurement of CP violation using time dilation to extend the lifetimes of B Mesons. This occurred because of the collider asymmetry. When two particles are colliding at different energies, it boosts the Lorentz factor of the decay. As we can see in the decay equation $L=\beta\gamma c\tau$ the Lorentz factor is one of the easiest variables we can change since it depends on 

$$\gamma = \frac{1}{\sqrt{1 - \frac{v^2}{c^2}}}$$

$v^2$ being the relative speed between lab frames. By creating the difference in energies it causes the resulting collision particle to not remain at rest in the lab frame, extending its lifetime. This will be an integral and necessary consideration for our tau collider. Since the tau only lives for $2.9 \times 10^{-13}$s we will need to find an optimal achievable gamma value that increases our tau's lifetime. Using asymmetry achieves these 2 neccessities: How long the beam survives, and an easier way to measure the result.

\subsubsection{\label{sec:tauz} Tau production through Z-Boson Decay}
 One of the greatest challenges in creating our tau collider is getting around the short lifetime. One suggested work-around solution is through a chain of multiple particle interactions. Since the tau particle is a very massive particle, it requires lots of energy to create. The main way to create tau particles is to create another particle that then decays into a tau anti-tau pair, or another combination of particles that contains a tau. Some examples of this is the Z boson, which produces a tau anti-tau pair approximately 3.369 percent of the time when it decays. \cite{pdg2022} Besides electron-positron annihilation there are other ways to generate a tau, W/Z and Higgz Boson decays. In the 1990's SLAC was able to produce millions of Z bosons and study them thoroughly, giving us plenty more data to work with than our tau lepton. Another possible choice of particle is the W boson, which produces a tau and tau neutrino 11.38 percent of the time, however, since the decay only produces one tau, it would need to occur twice in a row in order to produce a pair of taus for us to collide\cite{pdg2022}. This would bring the probability of this occurring to $.1138^2$, or $0.01295$ which is a 1.295 percent chance. The Higgs boson also produces a $\tau^+ \tau^-$ pair when it decays, with a 6 percent chance of occurring\cite{pdg2022}. However, the Higgs is much more difficult to create than a W or Z boson, as the production cross section of a Higgs boson is around 50 pico barns, while the Z boson is around 30 nanobarns\cite{pdg2022}. This means that the production of a Z boson is 600x more likely to occur than a Higgs production; Therefore, the only particle we can produce en masse, and with sufficient chance to decay into a $\tau^+ \tau^-$ pair is the Z boson. Since it is easier to create a Z boson than a $\tau^+ \tau^-$ pair, it is possible to create a frame boosted Z Boson using an asymmetric collider, that decays into a $\tau^+ \tau^-$ pair that is already moving at relativistic speeds.

\subsection{\label{sec:taucirc} Design of the collider}

\begin{figure}[H]
    \centering
    \includegraphics[width=0.5\textwidth]{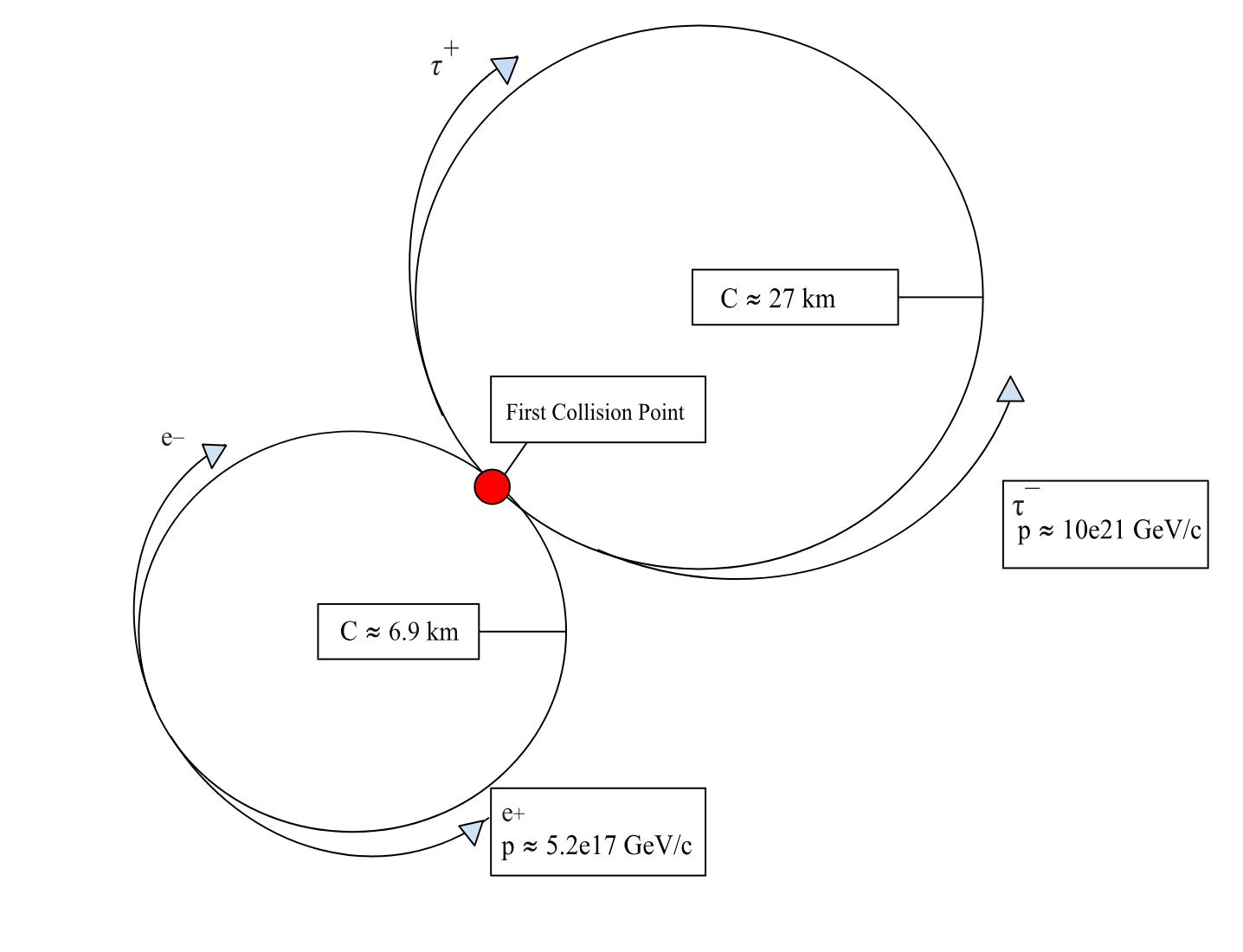}
    \caption{The proposed tau collider consists of two circular colliders, the smaller (left side ) being an electron positron collider while the larger (right side) collides $\tau$. Once the electrons and positrons collide to create $\tau^+$ and $\tau^-$, we can detect, focus, and accelerate taus using magnetic and electric fields (for which specific values are calculated in the sections above).}
    \label{fig:6}
\end{figure}

One approach to the construction would be to use the current LHC
and SPS tunnels to build our accelerator (Fig~\ref{fig:6}). However, 
the current technology used to create the electric and magnetic fields
necessary may not be up to the task. 

\subsection{\label{sec:kard} Potential engineering solutions: The Kardashev Scale}

The Kardashev scale was created as a way to rank civilizations' development, with the intent of classifying the amount of energy that can be utilized in ones environment. The three main types are, Type I uses all power available in one's planet, Type II which harnesses all power from one's star, and Type III controls all power from its entire galaxy. These scales were made with the idea that when a civilization reaches full power usage of its planet they are expected to reach out to their solar system and start harvesting energy from other planets. For reference, humanity is measured to be below Type I on the Kardashev scale~\cite{Kardashev_Scale}. 

Why is this important to the creation of a tau collider? The question we are looking at is there even enough possible energy on earth to make this happen, even before we can try to harness it all into one collider? It is estimated that earth has reached $2\times10^{13}$ watts which is still to the power of 4 times smaller than what is believed to be the maximum on earth(This puts our civilization at Type 0.7)~\cite{Planet_Energy}. The amount of watts needed to power our supposed tau collider would be an extremely large number but no where near as big as the amount of energy we have on planet earth. $2\times10^{13}$ J/s roughly converts to $1.2\times10^{32}$ eV. Hypothetically if we were able to find a way to channel a large amount of energy into our collider, we would be able to source it just from planet earth. 

A more recent advance in the Kardashev scale, which will be much more useful for our project was proposed by John Borrow\cite{Barrow_Scale}. Micro-dimensional categories of the Kardashev scale is going by the fact that humans have found it much more cost effective to manipulate their environment in much smaller dimensions rather than larger. This classification of types start at Type I-minus where civilization can manipulate objects on a normal scale like buildings and mining materials, all the way to Type $\Omega$-minus where civilization can manipulate the fundamental structure of space and time.

As of 2026, we are considered to be a Type III-minus civilization where we have mastered molecular engineering, created artificial material, and performed genetic engineering. With how quick our civilization has been moving through this scale, we might want to consider the microscale over the Kardashev. Are we closer to Type VI-minus, where we have mastery of the most elementary particles in our universe (quarks and leptons) or creating a collider that needs to be the size of the earth and harvest materials from other planets.~\cite{anti-Kardashev}. 


\section{\label{sec:mat} Materials Needed}
At the present moment, most high energy accelerator facilities such as the Large Hadron Collider and the RHIC utilize Niobium-Titanium magnets for beam focusing. While these magnets are very efficient at their job, they are very expensive to make and require a complicated manufacturing process. As we consider the components needed to operate a Tau accelerator the question of what materials are needed to build such a device, specifically with the focusing magnets and the RF cavities. Most current accelerators, such as the LHC, have used Niobium-Titanium (Nb-Ti) superconducting magnets that can create magnetic fields of a strength of 8.6 Teslas. However, in preparation for the new High-Luminosity LHC a set of Niobium-Tin magnets were created that can produce a field of ~11 Tesla. 

At the current stage of technological development this is the strongest mass-produced superconducting magnet that has ever been created for the context of a particle accelerator. Using this value of for magnetic field strength we can calculate the accelerator radius needed to reach the level of energy needed for the $\tau^{\pm}$ to live 1 millisecond. At 10 Zetta eV ($10^{21}$eV), the accelerator would need a radius of roughly $10^{13}$ meters which is approximately 66 AU. Firmly within the Oort cloud of the solar system it quickly become clear that there doesn't exist enough material on both the earth and likely a large number of asteroids to build an accelerator this large. Additionally, 
the $\tau$'s may not live long enough to make it around the collider
ring, perhaps necessitating multiple $\tau$ injection points.

On the other side of things, lets imagine a world in which we are space limited to a collider that must be built on earth. That means we are limited to a radius of around {$6 \times 10^{6}$} meters which would require a magnetic field strength of roughly $10^{11}$ Tesla. This is on a scale that we are nowhere near accomplishing with human-made technology and fields this strong have only been observed surrounding a neutron star which can have a field strength of ~{$10^{12}$} Tesla. Seeing as though we have not reached that point in our development that we could feasibly harness a neutron star and bend its magnetic field lines around the Earth for our collider, we would reason that this path is also an nonviable solution for the foreseeable future.

\section{\label{sec:cost} Cost Analysis}

With the incredible energy requirements needed for colliding $\tau$'s exceeding what is technological possible in the present day, it is a valid question to ask how much such a complex facility might cost to construct. For comparison lets look at the cost of the Tevatron located at Fermilab in the US and the LHC, the main accelerator at CERN in Geneva. Shown in the table below. are the construction cost, an inflation adjusted present day estimate, the energy levels achieved by the collider, and the circumference of the accelerator ring.

\begin{table}[h!] 
\label{Accelerator Cost Info}
\setlength{\tabcolsep}{2pt}
\begin{tabular}{c r r r r}  
\hline\hline
Collider  & Const. Cost  & Adj. Cost & Energy & Cost per Km  \\[0.5ex]
 & (USD) & (USD) & (GeV) & (USD/km) \\[0.5ex]

\hline 
Tevatron & 120 \cite{Tevatron_info} & 338 & 1000 & 65 \\   
LHC & 4600 \cite{LHC_Cost} & 6970 & 14000 & 230 \\ 
RHIC & 486.8 \cite{RHIC_info} & 950.0 & 500 & 250\\ [1ex] 
\end{tabular} 
\caption{Costs are listed in Millions of United States dollars. The construction cost (labeled on the table is Const. Cost) is the price to build that accelerator at the time of its construction. We accounted for inflation in Adj. Cost to see how much money we would need for the same buying power.} 
\label{table:nonlin}  
\end{table}

From these values we can estimate that the base cost for building an accelerator would be 65 - 250 million dollars per KM built. Due to the complex nature of a tau accelerator and the novelty of its design we estimate that the real cost per km would fall more closely to the 250 million USD mark. This is due to the extra depth needed to ensure adequate radiation shielding, the R\&D needed for new cooling systems, as well as the rising cost in materials and labor. We would like to note that this value also doesn't include the cost of the Detectors or the lab facilities that would need to be built for researchers.

From these values, should we need a $\tau$ collider that spanned the circumference of the Earth at the equator, we would need over 10 Trillion dollars which is about 1.3 times the US federal budget as of 2025. A project of this magnitude would have never been seen before in the history of the United States as well as the world. For comparison, at the peak of the Apollo program, NASA's spending only made up ~4\%\cite{apollo_funding_us} of the US federal budget.

\section{Summary}

In this paper we have argued for the scientific benefits of a
$\tau$ collider and mapped out the design parameters. There are
still engineering challenges to solve, however we anticipate that with
adequate funding, these challenges can be met before our sun 
evolves to a red giant stage and consumes the Earth. 

\appendix
\section{Motivation for this paper}

This paper grew out of the {\it PHYS 400: Nuclear and Particle Physics}
course at Siena University in the Spring 2026 term. The course is an
upper-level elective for Physics, Astrophysics, and Applied Physics
majors. The motivation was to provide the students some insight as
to how large-scale scientific experiments are developed. This was inspired
by course-based undergraduate research experiences (CUREs) developed
at other schools. Rather than try to do a full analysis of
real data, we focused
on the design process for a collider that is not possible with 
current or foreseeable technology. Given the that five (5) of the students
were Applied Physics, having a class project that incorporated some
level of engineering discussion was appropriate.

The first half of the term was spent
learning the basics of particle physics and accelerator
design parameters and the two weeks before this submission (April 1)
was spent crowdsourcing the research and writing of the sections 
of this paper among the twelve (12) students in the class.

For some students, English is not their first language and all of the
students are still learning to write scientific documentation. 
While the instructor went through
and cleaned up parts of the language, most of the text comes straight
from the students, even when it is not polished, 
so that they can see their own words in the paper.

Criticisms of the paper or the science should be directed at the 
instructor, Matt Bellis.

\bibliography{apssamp}

\end{document}